\documentclass[twocolumn,showpacs,superscriptaddress,amsmath,amssymb]{revtex4}
\usepackage{graphicx}
\usepackage{dcolumn}
\usepackage{bm}
\usepackage{bm}
\def\bs{\boldsymbol}

\def\sech{{\rm sech}\aa}          \def\csch{{\rm csch}\aa}
\def\Arg{{\rm Arg}\aa}            \def\Arctg{{\rm Arctg}\aa}
\def\arctg{{\rm arctg}\aa}        \def\vs{\vskip}
\def\tg{{\rm tg}\aa}              \def\ctg{{\rm ctg}\aa}
\def\sh{{\rm sh}\aa}              \def\ch{{\rm ch}\aa}
\def\th{{\rm th}\aa}              \def\tan{{\rm tan}\aa}
\def\cth{{\rm cth}\aa}            \def\f{\left}  \def\g{\right}

\def\grad{{\rm grad\hskip3pt}}    \def\div{{\rm div\hskip3pt}}
\def\huaD{\mathcal{D}}            \def\huaL{\mathcal{L}}
\def\Re{{\bf Re}\aa}              \def\Im{{\bf Im}\aa}

\def\etc{{\it etc.}}              \def\ie{{\it i.e. }}
\def\cf{{\it cf.}}                \def\eg{{\it e.g. }}
\def\cH{{\cal H}}                 \def\cV{{\cal V}}

\def\bd{\begin{document}}   \def\ed{ \end{document} }
\def\ie{{\it i.e.}\ }   \def\eg{{\it e.g.}\ }   \def\cf{{\it cf.}\ }
\def\etc{{\it etc.}\ }  \def\P{{\rm P}}
\def\bs{\boldsymbol} \def\nd{\noindent} \def\nbf{\nd\bf} \def\mn{\vskip0.5cm\nd}
\def\aa{\hskip3pt} \def\aaa{\hskip1.5pt}  \def\np{\newpage}
\def\md{\vskip0.3cm}
\def\bec{\begin{center}}    \def\eec{\end{center}}
\def\bct{\begin{center}}    \def\ect{\end{center}}  \def\cl{\centerline}
\def\hs{\hskip}  \def\vs{\vskip}  \def\lrw{\longrightarrow}
\def\bmp{\begin{minipage}}    \def\emp{\end{minipage}}
\def\beq{\begin{equation}}    \def\eeq{\end{equation}}
\def\bea{\begin{eqnarray}}    \def\eea{\end{eqnarray}}
\def\bes{\begin{eqnarray*}}    \def\ees{\end{eqnarray*}}
\def\bpm{\begin{pmatrix}} \def\epm{\end{pmatrix}}
\def\ben{\begin{enumerate}} \def\een{\end{enumerate}}
\def\btb{\begin{tabular}} \def\etb{\end{tabular}}
\def\btbb{\begin{tabbing}} \def\etbb{\end{tabbing}}
\def\af{\alpha} \def\bt{\beta}  \def\gm{\gamma}  \def\tr{{\rm tr}\,}
\def\lm{\lambda}  \def\Lm{\Lambda} \def\spic{^{\footnotesize(\rm S)}}
\def\hpic{^{\footnotesize(\rm H)}}   \def\upic{^{\footnotesize(\rm U)}}
\def\ipic{^{\footnotesize(\rm I)}}  \def\hn{\hat{\bs n}}
\def\sech{{\rm sech}\,}  \def\Arg{{\rm Arg}\,} \def\Arctg{{\rm Arctg}\,}
\def\arctg{{\rm arctg}\,}  \def\nbr{\nonumber} \def\dt{\delta}
 \def\Dt{\Delta} \def\ep{\epsilon} \def\ve{\varepsilon}
 \def\sm{\sigma}   \def\Sm{\Sigma}      \def\ta{\theta}  \def\Ta{\Theta}
 \def\om{\omega}   \def\Om{\Omega}    \def\kp{\kappa}  \def\gm{\gamma}
\def\tan{{\rm tan}\,}  \def\vf{\varphi}  \def\vt{\vartheta}
 \def\cH{{\cal H}}   \def\cV{{\cal V}}   \def\cD{{\cal D}\,}
 \def\cK{{\cal K}\,}   \def\Uvf{U_{\bs \vf}}
 \def\Gm{\Gamma}    \def\ih{\frac{i}{\hbar}}  \def\cI{{\cal I}}
 \def\tg{{\rm tg}\,}      \def\ctg{{\rm ctg}\,} \def\csch{{\rm csch}\,}
\def\sh{{\rm sh}\,}      \def\ch{{\rm ch}\,}   \def\th{{\rm th}\,}
\def\cth{{\rm cth}\,} \def\C{{\rm C}}   \def\grad{{\rm grad\hskip3pt}}
\def\div{{\rm div\hskip3pt}}  \def\L{{\rm L\hskip3pt}}
\def\lra{\longrightarrow}  \def\bone{{\bf 1}}
\def\qq{\qquad}   \def\fc{\frac}   \def\fnsz{\footnotesize}
\def\inint{\int_{-\infty}^{\infty}}  \def\ol{\overline}
 \def\ben{\begin{enumerate}} \def\een{\end{enumerate}}
\def\qd{\quad}  \def\qqd{\qquad} \def\btm{\begin{itemize}}
\def\etm{\end{itemize}}   \def\pl{\partial}  \def\huaN{\mathcal{N}}
\def\huaD{\mathcal{D}}  \def\huaL{\mathcal{L}} \def\d{{\rm d}}
\def\ddz{{\rm d}\over{{\rm d}z}}  \def\dv{{\rm d}} \def\huaG{\mathcal{G}}
\def\Re{{\bf Re}\;} \def\Im{{\bf Im}\;}  \def\g{\right}  \def\e{{\rm e}}
\def\f{\left} \def\r{\right} \def\la{\langle} \def\ra{\rangle}
\def\npg{\newpage} \def\ihb{\frac{i}{\hbar}}
\def\dbar{{\rm d}\hskip-5.6pt \rule[1.8mm]{2.0mm}{0.18mm}\hskip2pt}
\def\sdbar{{\rm d}\hskip-4.2pt \rule[1.45mm]{1.4mm}{0.12mm}\hskip2pt}
\def\ointcw{\mathop{\int\mkern-21.mu \circlearrowright}} 
\def\ointacw{\mathop{\int\mkern-20.mu \circlearrowleft}} 
\def\sointcw{\mathop{\int\mkern-19.mu \circlearrowright}} 
\def\sointacw{\mathop{\int\mkern-18.mu \circlearrowleft}} 
\def\ssointcw{\mathop{\int\mkern-18.5mu \circlearrowright}} 
\def\ssointacw{\mathop{\int\mkern-17.5mu \circlearrowleft}} 
\def\Solution{\vskip1mm\noindent {$\bs S\bs o\bs l\bs u\bs t\bs i\bs o\bs n$.}\quad}
\def\Res{{\bf Res}} \def\axiom{{\vskip2mm\noindent \bf Axiom\ }}
\def\bcs{\begin{cases}}  \def\ecs{\end{cases}}
\newcommand{\oiint}{\mathop{\makebox[-0,32em][l]
{$\bigcirc$}\int\!\!\!\!\!\int\makebox[-0.5em]{}}}

\begin{document}
\normalsize
\title{Pair distribution function of strongly coupled quark gluon plasma
in a  molecule-like aggregation model \footnote{supported by NSFC
under project No.10775056 and No.90503001.}}

\author{Yu Meiling\footnote{Email:\ yuml@iopp.ccnu.edu.cn}}

\affiliation{Department of Physics, Wuhan University, Wuhan 430072,
China}

\author{Xu Mingmei}

\affiliation{Institute of Particle Physics, Huazhong Normal
University, Wuhan 430079, China}

\author{Liu Zhengyou}

\affiliation{Department of Physics, Wuhan University, Wuhan 430072,
China}

\author{Liu Lianshou\footnote{Email:\ liuls@iopp.ccnu.edu.cn}}

\affiliation{Institute of Particle Physics, Huazhong Normal
University, Wuhan 430079, China}

\affiliation{Key  Laboratory of Quark \& Lepton Physics (Huazhong
Normal University), Ministry of Education, China}

\begin{abstract}
Pair distribution function for delocalized quarks in the  strongly
coupled quark gluon plasma (sQGP) as well as in the states at
intermediate stages of crossover from hadronic matter to sQGP are
calculated using a  molecule-like aggregation model. The shapes of
the obtained pair distribution functions exhibit the character of
liquid. The increasing correlation length in the process of
crossover indicates a diminishing viscosity of the fluid system.
\end{abstract}



\maketitle

\section{Introduction}

Relativistic heavy-ion collision experiments at RHIC have found
evidences for a new state of matter, the quark-gluon plasma
(QGP)~\cite{whitepapers1}\cite{whitepapers2}. The collective flow
phenomena, the transport properties and other theoretical
developments indicate that the new state of matter is not a weakly
coupled gas as expected, but a strongly coupled ideal fluid,
referred to as sQGP, probably in a wide temperature region
$T_c<T\lesssim 2T_c$~\cite{sQGP1}\cite{sQGP2}\cite{sQGP3}. What is
the microscopic dynamics responsible for the small viscosity of the
strongly coupled quark gluon plasma and how is the evolution of the
property of this fluid in the process of crossover from hadronic gas
to sQGP are still open questions.

Many attempts have been done to explain the small viscosity.
Theoretically, people obtain transport coefficient perturbatively by
using Kubo formula in the scope of thermal field
theory~\cite{kubo1}-\cite{kubo5} or solve the transport equation
from kinetic theory~\cite{kinetic1}-\cite{kinetic4}. They found that
it is not enough to account for the small viscosity if only
considering the perturbative interaction. While ref.~\cite{xuzhe}
claimed that they can reproduce the small viscosity for a gluon gas
if including $gg\to ggg$ bremsstrahlung in the pQCD calculation
within relativistic kinetic theory. Another way to reach small
viscosity is to solve the strongly coupled super-symmetric
Yang-Mills gauge theory by AdS/CFT
duality~\cite{adscft1}\cite{adscft2}. In phenomenology,
Ref.~\cite{shuyak} developed a new picture of QGP with multiple
colored bound states and the increased re-scattering among bound
states were expected to reduce the viscosity of QGP.

\begin{figure}
\includegraphics[width=0.5\linewidth]{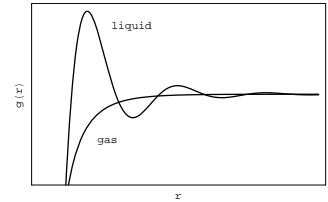}
\caption{\label{gr1} Sketch of the pair correlation function vs.
distance between two atoms in the gas and liquid phase.}
\end{figure}

Quantitative investigation on the liquid property of plasma is
possible by considering two particle correlation function in
coordinate space or its Fourier transform --- the structure function
in momentum space. The characteristic behavior of pair correlation
function reveals the intrinsic feature of plasma in different phase.
For example, in the case of liquid state the pair correlation
function exhibits a pronounced peak and one or two small and broad
additional peaks. The first peak corresponds in coordinate space to
the near-neighbor ``shell'' of atoms because of short-range order.
The next near-neighbor ``shell'' in the liquid is much less
prominent and the next outer shell may hardly visible due to the
lack of long-range order. In the case of a solid crystalline phase,
where a long-range order exists, a number of sharp peaks with
comb-like shape are observed. In the ideal gas phase, where no order
is present, the pair correlation function shows no clear structure.
A sketch of the pair correlation function for liquid and gas is
shown in Fig.~\ref{gr1}.

Though it is not easy to obtain the viscosity of a liquid
quantitatively from its pair correlation function, the tendency of
the viscosity could be inferred via the shape evolution of the
latter. As stated above the pair correlation function of liquid
exhibits a number of peaks with diminishing heights. The location of
the last visible peak can be taken as the range of effective
interaction in the liquid. The increasing of the latter results in a
diminishing mean free path, which in turn causing the reduction of
viscosity. Therefore, the larger distance the peaks of pair
distribution function with reduced amplitudes shows, the less
viscous the liquid is. Ref.~\cite{Thoma} calculated the correlation
function and structure function for QGP perturbtively by using hard
thermal loop approximation and it failed to represent the typical
character of liquid state since only weakly coupled QGP is
considered.

Recently, a dynamical percolation model based on the molecule-like
aggregation is proposed to describe the crossover transition between
hadronic matter and QGP~\cite{xumm}. In this model as the increase
of temperature, partons inside hadrons are delocalized to a certain
extent, tunneling between neighboring hadrons through bonds, to form
a sort of grape-shape QGP (gQGP), which is a special form of sQGP.
Inside the gQGP, partons move around with large cross section and
short mean free path, resulting in a quark-gluon matter possessing
the property of near-perfect fluid. In this article we investigate
the pair correlation function of the delocalized quarks from this
molecule-like aggregation model and discuss the information about
the structure of the grape shape QGP during the process of
crossover.

\section{\label{section2} A brief review on  molecule-like aggregation model}

In order to answer the question how to crossover from the hadronic
to partonic phase in QCD without contradiction with color
confinement, a basic assumption is proposed~\cite{xumm}, which
states that quarks in neighboring hadrons can be delocalized, i.e.
tunnel between the hadrons to form clusters. After delocalization
the clusters, which are molecule-like aggregations, are color
singlets and the original hadrons turned to colored objects, called
cells.

To exhibit the physical consequence of the assumption on {\it
molecule-like aggregation} a toy model basing on percolation
procedure is constructed~\cite{xumm}. In the simplified 2-D version
of the model the initial system is set to be a nucleon gas
consisting of $2\times 197$ cells, \ie hadrons, which are small
circles of hard-core radius $r_e=0.1$ fm distributed randomly in a
big circle of radius $R=7$ fm.

In calculating the probability for bond-formation between neiboring
cells, or delocalization of quarks, a non-relativistic model used in
nuclear force theory~\cite{wangfan} is utilized. The model
Hamiltonian for the 6 constituent quarks in the c.m. system of two
nearby cells is assumed to be
\begin{equation}
\mathcal{H}=\sum_{i=1}^{6}\left(m_{i}+\frac{p_{i}^{2}}{2m_{i}}\right)-T_{\rm
cm}+\sum_{i<j}V_{ij}^{C}, \label{eqH}
\end{equation}
where $T_{\rm cm}$ is the center-of-mass kinetic energy. When the
quarks $i,j$ belong to a same cell, a square-confinement potential
$V_{ij}^{C}=-a_{c}\vec{\lambda}_{i}\cdot\vec{\lambda}_{j}r_{ij}^{2}$
is assumed. When they belong to two nearby cells, the infinite
potential between them will drop down, forming a potential barrier,
and a parametrization $V_{ij}^{C}
=-a_{c}\vec{\lambda}_{i}\cdot\vec{\lambda}_{j}\frac{1-e^{-\mu
r_{ij}^{2}}}{\mu}$ is used, where $\mu$ is a model parameter,
proportional to temperature square from dimensional consideration.

In doing variational calculation for the ground state energy the
trial wave function of the two-cell system  in adiabatic
approximation is chosen to be an
antisymmetric six-quark product state 
\begin{equation}
    \f|\Psi_6(S)\g>=\mathcal {A}\f[\prod_{i=1}^{3}\psi_{\rm L}(\bs{r}_{i})
    \prod_{i=4}^{6}\psi_{\rm R}(\bs{r}_{i})\eta_{I_1 S_1}^{B_1}\eta_{I_2
    S_2}^{B_2}\chi_{c}^{B1}\chi_{c}^{B_2}\g]_{00},
\end{equation}
where $\mathcal {A}$ is the anti-symmetrization operator, which
permutes quarks between the two cells; $[\cdots]_{00}$ means that
the spin, isospin and color of the two cells are coupled to a
particular color singlet state with total spin and isospin equal
zero. For the orbital motion, we have the left (right) single-quark
orbital wave function $ \phi_{L}(\bs{r}_{i})=\f(\frac{1}{\pi
b^2}\g)^{\frac{3}{4}}\e^{-\frac{(\bs{r}_{i}+\frac{\bs{S}}{2})^2}{2b^2}},\
\phi_{R}(\bs{r}_{i})=\f(\frac{1}{\pi
b^2}\g)^{\frac{3}{4}}\e^{-\frac{(\bs{r}_{i}-\frac{\bs{S}}{2})^2}{2b^2}},$
where $\pm\frac{\bs{S}}{2}$ are the cell-centers, $S$ is the
distance between the two cells. $b$ is a baryon-size parameter.
Delocalized orbit is defined as
$\psi_{L}(\bs{r})=\frac{1}{N}[\phi_{L}+\epsilon\phi_{R}],\
  \psi_{R}(\bs{r})=\frac{1}{N}[\epsilon\phi_{L}+\phi_{R}],$
where $\epsilon$ is a variational parameter characterizing the
degree of delocalization and $N$ a normalization factor. At each
separation $S$, $\epsilon$ is determined by minimizing the energy
$E(S)=\frac{\f<\Psi_6(S)\f|\mathcal
{H}\g|\Psi_6(S)\g>}{\f<\Psi_6(S)|\Psi_6(S)\g>}.$ The model
parameters are: $m_{\rm u}=m_{\rm d}=313~{\rm MeV}$, $b=0.603~{\rm
fm}$, $a_{c}=101.14~{\rm MeV}/{\rm fm}^2$, while $\mu$ is leaving as
a free parameter.

\begin{figure}
\includegraphics[width=3.4in]{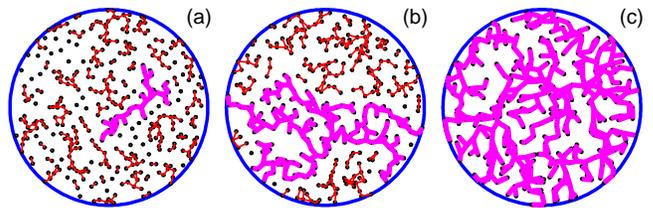}
\caption{\label{Fig. 2}  (Color online) Continuously distributed
cells connected by bonds form clusters. In (b) the big cluster
extending from the left- to the right-boundary is an {\it infinite
cluster}. In (c) all the cells are connected to an infinite
cluster.}
\end{figure}

It is found from variational calculation that a maximum distance
$S_0$ for delocalization exists for a fixed temperature, i.e. fixed
$\mu$. Using $S_0$ as input, a bond percolation model is
constructed. After the percolation procedure all the cells in the
system are grouped into clusters. In each cluster, any two cells are
connected by bonds with zigzag path, mimic the atoms in molecule,
cf. Fig.~\ref{Fig. 2}(a).

When temperature increases, the potential barrier between two cells
decreases, and the maximum distance $S_0$ for bond formation
increases. The crossover from hadronic phase to partonic phase
starts when an infinite cluster \ie a cluster extending from one end
of the big circle to the other end appears, cf. Fig.~\ref{Fig.
2}(b). The corresponding temperature is $T_c$. The colored partons
are confined in the group of cells connected by bonds, being able to
move around inside the big cluster via quantum tunneling through the
potential barriers, resulting in a quark-gluon matter with the
property of fluid inside the cluster. The crossover process is
completed when all the cells belong to one infinite cluster, cf.
Fig.~\ref{Fig. 2}(c), and the corresponding temperature is denoted
by $T_c'$ . In the following we will investigate the structure of
sQGP and of the intermediate states during the crossover from
hadronic phase to sQGP through the pair correlation function.

\section{\label{section3} Definition of pair correlation function}

\begin{figure}
\includegraphics[width=0.9\linewidth]{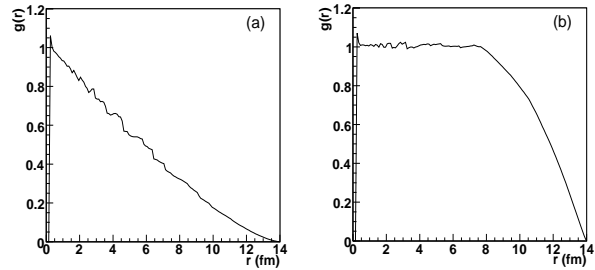}
\caption{\label{hadrongas} Pair correlation function for randomly
distributed hadron gas. (a) is without boundary correction and (b)
is with boundary correction. }
\end{figure}

In the liquid state theory, the pair correlation function $g(r)$ is
defined as the probability of finding two atoms in the liquid at a
distance $r$ from each other~\cite{gr1}\cite{gr2}. In two
dimensional space the quantity $2\pi\rho g(r)rdr$ is the mean number
of atoms inside a ring of radius $r$ with thickness $dr$, centered
on an ``average'' atom. In this expression, $\rho$ is the number
density of the bulk homogeneous liquid. From this definition we can
calculate $g(r)$ in two dimensional coordinate space in Monte Carlo
simulation by the formula:
\begin{equation}
g(r)=\frac{dN(r)}{2\pi\rho rdr}, \label{eqgr0}
\end{equation}
where $dN(r)$ is the number of atoms inside a ring with radius
$(r,r+dr)$ apart from the selected center atom. $dN(r)$ is
normalized by the uniform distribution, so that $g(r)=1$ when there
is no correlation. It needs to be noted that for a finite system,
the boundary effect has to be taken into account since it will
affect the normalization factor. Assume the selected central atom is
near the boundary, then part of the ring with radius from $r$ to
$r+dr$ may lie out of the finite system. Thus the normalization
factor should be $\theta\rho rdr$ instead of $2\pi\rho rdr$, where
$\theta$ is the angle of the arc inside the system. To eliminate the
boundary effect, we add a weight $w$ to the above formula, and
define
\begin{equation}
g(r)=\frac{dN(r)\cdot w}{2\pi\rho rdr}, \qquad w={2\pi\over\theta}.
\label{eqgr1}
\end{equation}
Fig.~\ref{hadrongas} is the pair correlation function $g(r)$ for a
randomly distributed hadron gas. Fig.~\ref{hadrongas}(a) shows the
result from Eq.~(\ref{eqgr0}) and Fig.~\ref{hadrongas}(b) is that
after the boundary effect correction by Eq.~(\ref{eqgr1}). Evidently
the correction is effective up to
the radius 7\;fm of the big circle. 
All the following calculations are restricted in this $r$ range.

To calculate the pair correlation function for quarks in the
above-mentioned  molecule-like aggregation model, the usual
definition about $g(r)$ has to be modified since, due to color
confinement the correlation between quarks does not happen along
straight lines in geometrical space but are along the zigzag path of
the connecting bonds. In order to reveal the structure information
of the sQGP inside clusters, the correlation function must be
defined as a function of the zigzag distance $D$ along bonds between
quarks, \ie $g(D)$. The zigzag distance $D$ is referred to as
``chemical distance'' in the language of percolation
theory~\cite{chemicaldistance}. The formula to calculate $g(D)$ in 2
dimensional space is
\begin{equation}
g(D)=\sum_r\frac{dN(D,r)\cdot w}{2\pi\rho rdr}. \label{eqgd}
\end{equation}
In the calculation, a quark is randomly selected as center and the
distance $D$ and $r$ between this center and any other quarks in the
same cluster are evaluated. $dN(D,r)$ is the number of quarks in the
small region $(D,D+dD;r,r+dr)$. The normalization $2\pi \ta r\d r$
for normal liquid is still used. The color confinement in the
present case, which causes quarks to be able to communicate only
through tunneling bonds, makes the resulting $g(D)$ tends to a value
smaller than unity.

\begin{figure}
\includegraphics[width=0.9\linewidth]{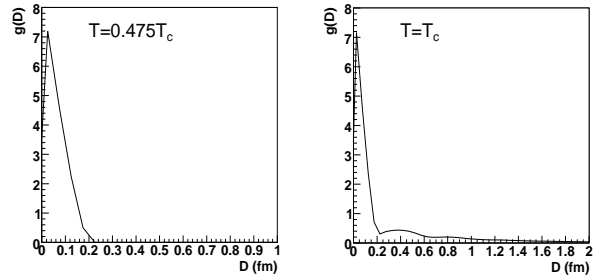}
\caption{\label{s019} Quark pair correlation function at temperature
$T=0.475T_c$ and $T=T_c$ in  molecule-like aggregation model.}
\end{figure}

\section{\label{section4} pair correlation function from  molecule-like aggregation model}

In our  molecule-like aggregation model, the formation of infinite
clusters is taken as the appearance of a new constituent \ie sQGP in
the system. The temperature where infinite clusters start to form is
referred to as $T_c$ and all the temperatures are scaled by it while
investigating the quark pair correlation function inside clusters.

\begin{figure*}
\includegraphics[width=0.7\linewidth]{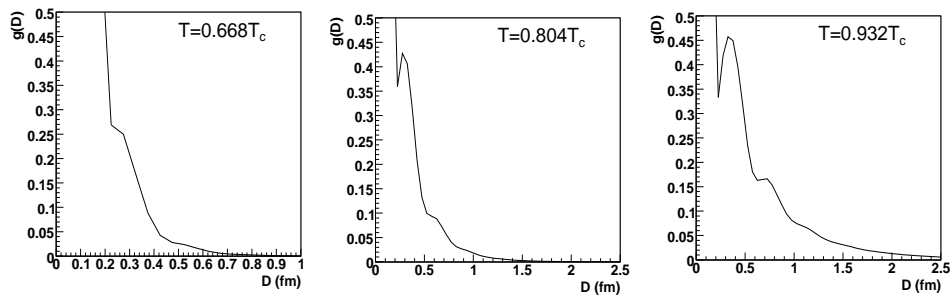}
\caption{\label{beforeTc} Quark pair correlation function $g(D)$ in
the range $D>0.2$ fm before $T_c$ in the molecule-like aggregation
model.}
\end{figure*}
\begin{figure}
\includegraphics[width=0.7\linewidth]{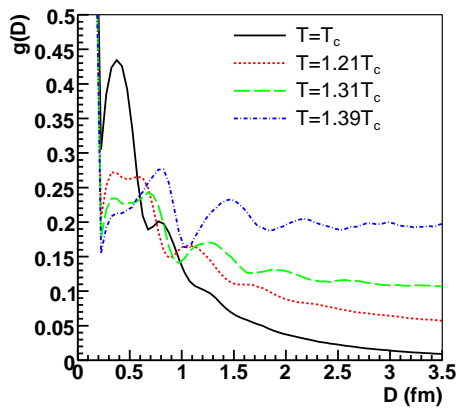}
\caption{\label{afterTc} (color online)  Quark pair correlation
function $g(D)$ in the range $D>0.2$ fm from the start of crossover
to the end of crossover in the  molecule-like aggregation model.}
\end{figure}

At each temperature, one quark is randomly selected from a randomly
chosen cluster in the system as the center particle. The geometrical
distance $r$ and chemical distance $D$ between this center quark and
any other quarks inside this cluster are calculated. Pair
correlation function $g(D)$ is then evaluated by Eq.~(\ref{eqgd}).
Fig.~\ref{s019} shows the pair correlation function for $T=0.475
T_c$ and $T=T_c$. At $T=0.475 T_c$, the maximum bond length
$S_0=0.19$\;fm is less than the smallest distance between any two
cells \ie $S_0<2 r_e$ and there is no bond formed between different
cells. Thus the peak at $D<0.2$\;fm in Fig.~\ref{s019} shows the
intra-cell structure, which is analog to the unpenetrable distance
in the usual liquid state theory. When $T=T_c$, infinite clusters
appear with small probability, and we see from the right panel of
Fig.~\ref{s019} that small bumps are shown at larger $D$ beside the
peak at $D<0.2$\;fm. To study the quark correlation induced by
delocalization and the formation of bonds, we will not look at the
trivial intra-cell correlation and only care the region where the
new correlations appear. In the following figures, only $g(D)$ in
the range $D>0.2$ fm is drawn to focus on the structure of the pair
correlation function induced by quark delocalization.

Fig.'s~\ref{beforeTc} show the $g(D)$ evolution at $T<T_c$. When $T$
is much less than $T_c$, i.e. long before crossover started, there
are no clear correlation peaks, while when going near to crossover
temperature, shoulder appears and quickly grows into the first
pronounced peak at $T=0.804 T_c$. As $T$ increases further new
shoulder emerges following this peak and the second peak forms when
$T=T_c$, cf. the full line in Fig.~\ref{afterTc}.

From the evolution of the pair correlation function with
temperature, we can get the following information about the
structure of gQGP formed in the molecule-like aggregation model:

(1) The first highest peak of the pair correlation function
represents the correlation from the nearest neighboring quarks
between different hadrons. The other small peaks are diminishing as
$D$ increases because of the long range disorder. The shape of
$g(D)$ around $T_c$ shows the typical short-range-order behavior of
liquid state, which indicates that the quark matter in the
molecule-like aggregation model possesses liquid structure.

(2) During the process of crossover, \ie $T_c<T<1.39T_c$, the
position of peaks shift to larger $D$ as the temperature increases,
cf. Fig.~\ref{afterTc}, showing that the correlation length
increases with temperature. This indicates that the viscosity is
getting smaller and smaller when temperature increases.

\section{Conclusion}

The molecule-like aggregation model provides a clear picture for the
structure of the matter formed in crossover, \ie gQGP, which is a
special form of sQGP, and the evolution of the structure in the
process of crossover. In this picture, quarks are moving from one
cell, which is initially hadron, to the other through bonds to form
grape shape quark matter. Basing on this model the pair distribution
function related to chemical distance $D$ for gQGP and its evolution
in the whole process of crossover are investigated. The typical
behavior of pair distribution function demonstrates that liquid
structure exists for the gQGP. The temperature dependence of the
correlation range qualitatively shows that the viscosity is getting
lower as the temperature increases during the crossover process.

It is worth while noticing that the above conclusion on the behavior
of pair correlation function in the process of crossover is mainly
based on the molecule-like aggregation assumption itself, and does
not rely much on the toy model used in the calculation. It could be
expected that the qualitative features shown in Fig's.2 and 6 will
persist in a more realistic case.

\end{document}